\documentstyle[prd,tighten,aps]{revtex}

\begin{document}
\draft
\date{\today}

\twocolumn[\hsize\textwidth\columnwidth\hsize\csname
@twocolumnfalse\endcsname

\setlength{\textwidth}{16truecm}
\setlength{\textheight}{22truecm}
\setlength{\topmargin}{-1truecm}
\setlength{\oddsidemargin}{0truecm}
\newcommand{\p}{\partial}

\title{New remarks on the linear constraint self-dual boson and Wess-Zumino terms}
\author{Everton M. C. Abreu and Alvaro de Souza Dutra}
\address{Departamento de F\'\i sica e Qu\'\i mica, Universidade Estadual Paulista, \\
Av. Ariberto Pereira da Cunha 333, Guaratinguet\'a, 12516-410, S\~ao Paulo,
SP, Brazil, \\
E-mail: everton@feg.unesp.br and dutra@feg.unesp.br}

\maketitle

\begin{abstract}
In this work we prove in a precise way that the soldering formalism can be applied to 
the Srivastava chiral boson (SCB), in contradiction with some results appearing in the literature.
We have promoted a canonical transformation that shows directly that the SCB is composed of two Floreanini-Jackiw's particles with the same chirality which spectrum is a vacuum-like one.
As another conflictive result we have proved that a Wess-Zumino term used in the literature consists of the scalar field,
once again denying the assertion that the WZ term adds a new degree of freedom to the SCB theory in order to modify the physics of the system.  
\end{abstract}


\vskip2pc]

\section{Introduction}

The research in chiral bosonization has begun many years back with the
seminal paper of W. Siegel \cite{siegel}. Floreanini-Jackiw have offered
later some different solutions to the problem of a single self-dual field 
\cite{fj} proposing a non-anomalous model. The study of chiral bosons has
blossomed thanks to the advances in some string theories \cite{ghmr} and in
the construction of interesting theoretical models \cite{grs}. They also
play an important role in the studies of the quantum Hall effect \cite{ws}.
The introduction of a soliton field as a charge-creating field obeying one
additional equation of motion leads to a bosonization rule \cite{ggkrs}.
Stone \cite{ms} has shown that the method of coadjoint orbit \cite{orbitas},
when applied to a representation of a group associated with a single affine
Kac-Moody algebra, generates an action for the chiral WZW model \cite{wzw},
a non-Abelian generalization of the Floreanini and Jackiw (FJ) model.

A self-dual field in two dimensions is a scalar field which satisfies the
self-dual constraint (self-dual condition) $(\eta^{\mu\nu}+\epsilon^{\mu%
\nu})\partial_{\nu}\phi=0$ or $\dot{\phi}=\phi^{\prime}$, where a dot means
time derivation and prime, space derivation. In the formulation of
Floreanini and Jackiw \cite{fj}, the space derivative of the field instead
of the field itself satisfies the self-dual condition, i.e., $(\partial_0 -
\partial_1)\partial_1 \phi=0$, and the field violates the microcausality
postulate \cite{ggrs2}.

Trying to overcome these difficulties, Srivastava \cite{srivastava}
introduced an auxiliary vector field $\lambda_{\mu}$ coupled with a linear
constraint and constructed a Lorentz-invariant Lagrangian for a scalar
self-dual field. Although Harada \cite{harada} and Girotti {\it et al} \cite
{ggr} have pointed out consistency problems with the Srivastava model at the
quantum level, the linear formulation strictly describes a chiral boson from
the point of view of equations of motion at the classical level. Some
methods were used to quantize the theory \cite{quantization}. The extension
to $D=6$ was accomplished in \cite{mm} as well as its supersymmetric case 
\cite{deri}.

On the other hand, the concept of soldering \cite{ms,harada2} has proved
extremely useful in different contexts. The soldering formalism essentially
combines two distinct Lagrangians manifesting dual aspects of some symmetry
to yield a new Lagrangian which is destituted of, or rather hides, that
symmetry. The quantum interference effects, whether constructive or
destructive, among the dual aspects of symmetry, are thereby captured
through this mechanism \cite{abw}. The formalism introduced by Stone could
be interpreted recently as a new method of dynamical mass generation \cite
{abw}. This technique parallels a similar phenomenon in two dimensional
field theory known as Schwinger mechanism \cite{schw} that results from the
interference between right and left massless self-dual modes of chiral
Schwinger model \cite{jr} of opposite chiralities \cite{abw}.

Furthermore, an important ingredient in the study of such kind of systems
are the so called Wess-Zumino (WZ) terms \cite{wz}, which are introduced in
the theory in order to recover the gauge invariance \cite{fs}. In \cite
{miao2}, it was proposed a new way of the derivation of the WZ counterterm.
It was based on the generalized Hamiltonian formalism of Batalin and Fradkin 
\cite{bf} which have suggested a kind of quantization procedure for
second-class constraint systems to which anomalous gauge theory belong \cite
{fs,faddeevkss}. The final action obtained, dependent on an arbitrary
parameter, has been constructed in order to become the Srivastava model
gauge invariant. The Lorentz invariance requirement has fixed the parameter
in two possible values which generates two possible WZ terms. The result,
with one of the WZ terms, after a kind of chiral decomposition, was that the
SCB spectrum is composed of two opposite FJ's chiral bosons, similarly to
what happens with the Minimal Chiral Schwinger Model \cite{dutra} . The
conclusion, however, was that the WZ term so obtained have {\it added} a new
physical degree of freedom, an antichiral boson, to the spectrum and
therefore changes the self-dual field into a massless scalar. Besides, in
another similar paper, Miao and Chen \cite{miao} have asserted that it is
impossible to apply the soldering formalism \cite{ms,harada2} to solder two
opposite chiral aspects of the model proposed by Srivastava, as was
successfully accomplished in the Siegel and Floreanini-Jackiw theories \cite
{adw}. It was pointed out that the method was invalid in the linear
formulation because of the inequivalence of Srivastava's and Siegel and
FJ's. Hence, to promote the fusion, it was constructed a chiral counterterm 
\cite{miao2} for the linear formulation of the chiral bosons. This
counterterm was the same Wess-Zumino term mentioned above.

In this work we have demonstrated that both conclusions are not really true.
We have applied successfully the soldering formalism and showed that the
interference on-shell of two SCB results in a massless scalar field. As
another result, we have performed essentially a canonical transformation
(CT) \cite{djt,bg} (as a special case of CT, we have used the dynamical
decomposition \cite{aw}, which promotes a separation of a chiral theory in
its dynamical and symmetry parts) and the outcome showed, in an exact way,
that the spectrum is already composed of two FJ's chiral bosons with the
same chirality confirming the well known result that the SCB has two degrees
of freedom thanks to the linear constraint structure \cite{bazeia}. Besides,
we have showed that the WZ term introduced in \cite{miao2} is in fact a
scalar field, i.e., it is composed of two FJ with opposite chiralities. So,
it is obvious that the WZ terms introduced naturally these particles since
the spectrum of the SCB is a vacuum-like one\cite{clo}.

In section {\bf 2} we have made a short review of the soldering formalism.
In section {\bf 3}, it was carried out the soldering of two SCB's models.
The dynamical decomposition of the theory and a discussion of the WZ term
were accomplished in section {\bf 4}. The conclusions are depicted in
section {\bf 5}.

\section{The soldering formalism}

In this section we will follow basically the reference \cite{w2} to make a
short review of the method of soldering two opposite chiral versions of a
given theory. For more details, the interested reader can see \cite
{abw,ainw,ad}.

The basic idea of the soldering procedure is to raise a global Noether
symmetry of the self and anti-self dual constituents into a local one, but
for an effective composite system, consisting of the dual components and an
interference term.

An iterative Noether procedure was adopted \cite{w2} to lift the global
symmetries. Therefore, assume that the symmetries in question are being
described by the local actions $S_{\pm}(\phi_{\pm}^\eta)$, invariant under a
global multi-parametric transformation 
\begin{equation}  \label{ii10}
\delta \phi_{\pm}^\eta = \alpha^\eta\;\;,
\end{equation}
where $\eta$ represents the tensorial character of the basic fields in the
dual actions $S_{\pm}$ and, for notational simplicity, will be dropped from
now on. As it is well known, we can write, 
\begin{equation}
\delta S_{\pm}\,=\,J^{\pm}\,\partial_{\pm}\,\alpha\;\;,
\end{equation}
where $J^{\pm}$ are the Noether currents.

Now, under local transformations these actions will not remain invariant,
and Noether counter-terms become necessary to reestablish the invariance,
along with appropriate auxiliary fields $B^{(N)}$, the so-called soldering
fields which has no dynamics. Nevertheless we can say that $B^{(N)}$ is an
auxiliary field which makes a wider range of gauge-fixing conditions
available \cite{harada2}. In this way, the $N$-action can be written as, 
\begin{equation}  \label{ii20}
S_{\pm}(\phi_{\pm})^{(0)}\rightarrow S_{\pm}(\phi_{\pm})^{(N)}=
S_{\pm}(\phi_{\pm})^{(N-1)}- B^{(N)} J_{\pm}^{(N)}\;\; .  \nonumber
\end{equation}
Here $J_{\pm}^{(N)}$ are the $N-$iteration Noether currents. For the self
and anti-self dual systems we have in mind that this iterative gauging
procedure is (intentionally) constructed not to produce invariant actions
for any finite number of steps. However, if after N repetitions, the non
invariant piece end up being only dependent on the gauging parameters, but
not on the original fields, there will exist the possibility of mutual
cancelation if both self and anti-self gauged systems are put together.
Then, suppose that after N repetitions we arrive at the following
simultaneous conditions, 
\begin{equation}  \label{ii30}
\delta S_{\pm}(\phi_{\pm})^{(N)} \neq 0 \qquad \mbox{and} \qquad \delta
S_{B}(\phi_{\pm})=0\;\;,
\end{equation}
with $S_B$ being the so-called soldered action 
\begin{equation}  \label{ii40}
S_{B}(\phi_{\pm})=S_{+}^{(N)}(\phi_{+}) + S_{-}^{(N)}(\phi_{-})+ %
\mbox{Contact Terms}\;\;,
\end{equation}
where the Contact Terms are generally quadratic functions of the soldering
fields. Then we can immediately identify the (soldering) interference term
as, 
\begin{equation}  \label{ii50}
S_{int}=\mbox{Contact Terms}-\sum_{N}B^{(N)} J_{\pm}^{(N)}\;\;.
\end{equation}
Incidentally, these auxiliary fields $B^{(N)}$ may be eliminated, for
instance, through its equations of motion, from the resulting effective
action, in favor of the physically relevant degrees of freedom. It is
important to notice that after the elimination of the soldering fields, the
resulting effective action will not depend on either self or anti-self dual
fields $\phi_{\pm}$ but only in some collective field, say $\Phi$, defined
in terms of the original ones in a (Noether) invariant way 
\begin{equation}
S_{B}(\phi_{\pm })\rightarrow S_{eff}(\Phi )\;\;.  \label{ii60}
\end{equation}
Analyzing in terms of the classical degrees of freedom, it is obvious that
we have now a bigger theory. Once such effective action has been
established, the physical consequences of the soldering are readily obtained
by simple inspection.

\section{The soldering of two Srivastava's self-dual bosons}

The Srivastava action for a left-moving chiral boson, is 
\begin{equation}  \label{01}
{\cal L}^{(0)}_{\phi} = \partial_+\phi\partial_-\phi\,+\,\lambda_{+}
\partial_-\phi\;\;,
\end{equation}
where we have used the light-front variables $\partial_\pm = {\frac{1 }{%
\sqrt{2}}} (\partial_0 \pm \partial_1)$ and $\lambda_{\pm}=\lambda_0 \pm
\lambda_1$.

Following the steps of the soldering formalism studied in the last section,
we can start considering the variation of the Lagrangians under the
transformations, 
$\delta\,\phi\,=\,\alpha$ and $\delta\,\lambda_{+}\,=\,0$. 
We will write only the main steps of the procedure.

In terms of the Noether currents we can construct 
\begin{equation}
\delta {\cal L}^{(0)}_{\phi}\,=\,J^{\mu}_{\phi}\,\partial_{\mu}\,\alpha\;\;,
\end{equation}
where $\mu=+,-$, $J^{+}_{\phi}\,=\,0$ and $J^{-}_{\phi}\,=\,2\,\partial_+\,%
\phi\,+\,\lambda_{+}$. 

The next iteration, as seen in the last section, can be performed
introducing auxiliary fields, the so-called soldering fields 
\begin{equation}
{\cal L}^{(1)}_{\phi}\,=\,{\cal L}^{(0)}_{\phi}\,-\,B_{\mu}\,J^{\mu}_{\phi}%
\;\;,
\end{equation}

\noindent and one can easily see that the gauge variation of ${\cal L}%
^{(1)}_{\phi}$ is

\begin{equation}  \label{34}
\delta {\cal L}^{(1)}_{\phi}\,=\,-\,2\,B_{-}\,\delta\,B_{+}\;\;,
\end{equation}
where we have defined the variation of $B_{\pm}$ as $\delta
B_{\pm}=\partial_{\pm}\alpha$, and we see that the variation of ${\cal L}%
^{(1)}_{\phi}$ does not depend on $\phi$. It is the signal to begin the
process with the other chirality, which is given by 
\begin{equation}
{\cal L}^{(0)}_{\rho} = \partial_+\rho\partial_-\rho\,+\,\lambda_{-}
\partial_+\rho\;\;,
\end{equation}
and again, let us construct the basic transformations $\delta\,\rho\,=\,%
\alpha$ and $\delta\,\lambda_{-}\,=\,0$.

The Noether\'{}s currents are $J^{+}_{\rho}\,=\,2\,\partial_+\,\rho\,+\,%
\lambda_{-}$ and $J^{-}_{\rho}\,=\,0$ 
and the variation of the final iteration is $\delta {\cal L}%
^{(1)}_{\rho}\,=\,-\,2\,B_{-}\,\delta\,B_{+}$.

Now we can see that the variation of ${\cal L}^{(1)}_{\phi,\rho}$ does not
depend neither on $\phi$ nor $\rho$. Hence, as explained before, we can
construct the final (soldered) Lagrangian as 
\begin{eqnarray}  \label{35}
{\cal L}_{TOT}&=&\,{\cal L}_{L}\,\oplus\,{\cal L}_{R}  \nonumber \\
&=&{\cal L}^{(1)}_{\phi}\,+\,{\cal L}^{(1)}_{\rho}\,+\,B_+\,B_-  \nonumber \\
&=&{\cal L}^{(0)}_{\phi}\,+\,{\cal L}^{(0)}_{\rho}\,-\,B_+\,J^+\,-\,B_-\,J^-
\,+\,B_+\,B_-
\end{eqnarray}
which remains invariant under the combined symmetry transformations for $%
(\phi,\rho)$ and $(\lambda_{+},\lambda_{-})$, i.e., $\delta\,{\cal L}%
_{TOT}\,=\,0$.

Following the steps of the algorithm depicted in the last section, we have
to eliminate the soldering fields solving their equations of motion which
result in $B_{\pm}\,=\,J^{\mp}$ where $J^{\pm}=J^{\phi,\rho}$.

Substituting it back in (\ref{35}) we have the final effective Lagrangian
density 
\begin{eqnarray}  \label{37}
& &{\cal L}_{TOT}=\,(\,\partial_-\,\phi\,-\,\partial_-\,\rho\,)\,
(\,\partial_+\,\phi\,-\,\partial_+\,\rho\,)\,  \nonumber \\
&+&\lambda_{+}\,(\,\partial_-\,\phi-\partial_-\,\rho\,)
-\lambda_{-}\,(\,\partial_+\,\phi-\partial_+\,\rho\,) -{\frac{1}{2}}%
\,\lambda_{+}\,\lambda_{-}  \nonumber \\
&=&\,\partial_-\,\Phi\,\partial_+\,\Phi+\lambda_{+}\,\partial_-\,\Phi -
\lambda_{-}\,\partial_+\,\Phi\, -{\frac{1}{2}}\,\lambda_{+}\,\lambda_{-}\;\;.
\end{eqnarray}
where the new compound field are defined as $\Phi=\phi\,-\,\rho$.

As we can see we have a second order term in the Lagrange multipliers.
Solving the equations of motion for the multipliers, we obtain that, 
\begin{eqnarray}  \label{24}
\lambda_{-}\,=\,2\,\partial_-\,\Phi \qquad \mbox{and} \qquad
\lambda_{+}\,=\,-\,2\,\partial_+\,\Phi\;\;.
\end{eqnarray}

Substituting the equations (\ref{24}) in (\ref{37}) we have 
\begin{eqnarray}
{\cal L}_{TOT}\,=\,-\,{\frac{1}{2}}\,\partial_{\mu}\,\Phi\,\partial^{\mu}\,%
\Phi,
\end{eqnarray}
which represents the massless scalar field action.

Hence, we have demonstrated in a precise way that it is possible to use the
soldering formalism to promote the fusion of two opposite SCB, in
contradiction with the assertion done in \cite{miao}. Finally, one can
conclude that, starting from these inconsistent Lagrangian densities, it is
recovered, in the soldering procedure, a consistent model which is, in fact,
the free scalar field. However, this result was not the expected one, but we
will come back to this issue later.

In the next section we will investigate the spectrum of the Srivastava model
constructing a canonical transformation \cite{bg}, i.e., using the special
case of the dynamical decomposition \cite{aw}. The objective is to analyze
the result obtained by Miao {\it et al} \cite{miao2} previously with the
alternative construction of the Wess-Zumino term of the Srivastava theory.

\section{The dynamical decomposition of the Srivastava model}

In the Hamiltonian formulation, canonical transformations can be sometimes
used to decompose a composite Hamiltonian into two distinct pieces. A
familiar example \cite{djt}, is the decomposition of the Hamiltonian of a
particle in two dimensions moving in a constant magnetic field and quadratic
potential. It can be shown that this Hamiltonian can be separated into two
pieces corresponding to the Hamiltonians of two one dimensional oscillators
rotating in a clockwise and a anti-clockwise directions, respectively. Let
us now make a canonical transformation analysis of the SCB. In this case,
that the theory is already a chiral one, we will promote a dynamical
decomposition of it, i.e., the theory will be decomposed in its dynamical
and symmetry parts. If the theory is not invariant, the result will show
only the dynamics of the system. To perform this we have to make a canonical
transformation \cite{bg} in (\ref{01}) using the Faddeev-Jackiw first-order
procedure.

At this point, some interesting comments are in order. The inconsistencies
of the SCB model at the quantum level, discussed in some works \cite
{harada,ggr}, can be verified from another point of view. This is done by
comparing the Lagrangian density of the SCB in Minkovisky space, i.e., 
\begin{eqnarray}  \label{llag}
{\cal L} &=&{\frac{1}{2}}\,\partial _{\mu }\phi \,\partial ^{\mu }\phi
\,+\,\lambda _{\mu }\,(g^{\mu \nu }\,-\,\epsilon ^{\mu \nu }\,)\,\partial
_{\nu }\,\phi  \nonumber \\
&=&{\frac{1}{2}}\,(\,\dot{\phi}^{2}\,-\,{\phi ^{\prime }}^{2}\,)\,+\,\lambda
\,(\,\dot{\phi}\,-\,{\phi ^{\prime }}\,)
\end{eqnarray}
where $\lambda = \lambda _{+}$, with that of the bosonized version of the
CSM 
\begin{eqnarray}
{\cal L}\,&=&\,{\frac{1}{2}}\,(\partial _{\mu })^{2}\,+\,e\,(\,g^{\mu \nu
}\,-\,\epsilon ^{\mu \nu }\,)\,\partial _{\mu }\,\phi \,A_{\nu }\,  \nonumber
\\
&+&\,\frac{a\,e^{2}}{2}\,A_{\mu }^{2}\,-\,{\frac{1}{4}}\,F_{\mu \nu }^{2}
\end{eqnarray}
and to note that the former is in fact a particular case of the latter,
where one should take care of the identifications: $a=0$ and $A_{\mu }\to
\lambda _{\mu }$, an external field with vanishing field strength. Now, one
can relate the inconsistency of the SCB with that of the CSM with the
regularization ambiguity parameter $a=0$, as shown by Girotti {\it et al} 
\cite{giro}. Now let us recover the discussion on the SCB, by doing its
dynamical decomposition and then discussing how and why the WZ terms
introduced in \cite{miao2} recover its quantum consistency.

The canonical momentum is defined by $\pi \,=\,\dot{\phi}\,+\,\lambda ,$ and
substituting it back in (\ref{llag}) to obtain the first-order form we have 
\begin{equation}  \label{02}
{\cal L}=\pi \,\dot{\phi}-{\frac{1}{2}}\,\pi ^{2}+\pi \,\lambda \,-\,{\frac{1%
}{2}}\,\lambda ^{2}-{\frac{1}{2}}\,{\phi ^{\prime }}^{2}-\lambda \,{\phi }%
^{\prime }\;\;,
\end{equation}

Now, as we have mentioned before, we have to do the following canonical
transformation 
\begin{equation}  \label{05}
\phi=\eta\,+\,\sigma \qquad \mbox{and} \qquad
\pi=\eta^{\prime}\,-\,\sigma^{\prime}\;\;,
\end{equation}
which is defined as a dynamical decomposition. Notice that $\phi$ is a
chiral field already. So, in this way, this canonical transformation will
allow us to know exactly what is the Srivastava chiral boson. Hence,
substituting (\ref{05}) in (\ref{02}) we have as a result 
\begin{eqnarray}
{\cal L}_{DD}={\eta^{\prime}}\,\dot{\eta} - {\eta^{\prime}}^2 -{%
\sigma^{\prime}} \,\dot{\sigma} - {\sigma^{\prime}}^2\, - 2\,\lambda\,{%
\sigma^{\prime}} - {\frac{1 }{2}}\,\lambda^2\;\;.  \nonumber
\end{eqnarray}
Again, solving the equations of motion for the $\lambda$-field we have $%
\lambda\,=\,-\,2\,{\sigma^{\prime}}$, and, substituting back, 
\begin{eqnarray}  \label{final}
{\cal L}_{DD}\,=\,{\eta^{\prime}}\,\dot{\eta}\,-\,{\eta^{\prime}}^2\,-\,({%
\sigma^{\prime}}\,\dot{\sigma}\,-\,{\sigma^{\prime}}^2)\;\;.
\end{eqnarray}

We can see clearly that this action represents two Floreanini-Jackiw's (FJ)
chiral bosons. Each one with the same chirality. This is caused by the fact
that the Lagrange multiplier has acquired dynamics because of the linear
constraint form. In fact, we are demonstrating that (\ref{llag}) has two
degrees of freedom, represented in (\ref{final}) by $\eta$ and $\sigma$,
differently from Siegel's approach, where $\lambda$ is a pure gauge degree
of freedom. This result corroborates the one found by Bazeia in\cite{bazeia}
analyzing the linear constraint chiral boson quantum mechanics. We can say
that both particles in (\ref{final}) act like a Gupta-Bleuler's pair so that
each chiral excitation destroys the other and the Hilbert space is composed
of vacuum. This result confirms the one found in \cite{clo}.

Hence, in the soldering process of the SCB, each FJ's chiral boson interact
with its opposite chiral partner, so that the final result represents a
scalar field. We can observe also that the linear constraint formulation of
the chiral boson does not contain the Hull noton \cite{hull}, a nonmover
field that cancels out the anomaly of the Siegel model (an alternative
fermionic noton was introduced in \cite{dgr}), which is expected since the
SCB is not gauge invariant.

The result (\ref{final}) contradicts the result obtained in \cite{miao2} in
the following way. There, firstly it was built a final action composed of
the Srivastava action plus a WZ term with an arbitrary parameter. The
Lorentz invariance fixed the parameter in two possible values which
originated two different WZ terms. Hence, one of the actions obtained, after
a kind of chiral decomposition, is shown to have two FJ's particles of
opposite chiralities, in an analogous fashion to what happens with the usual
CSM \cite{dutra92}. Besides, the final Lagrangian obtained contain the BF
fields \cite{bf} used to construct the WZ term \cite{fik}. The conclusion
was that the WZ term constructed have added a new degree of freedom to the
theory in the form of an antichiral boson. 
On the other hand, we can see what is really happening through a careful
analysis of the two WZ terms introduced in \cite{miao2}. It is not difficult
to see that the first WZ term defined in \cite{miao2}, i.e. , 
\begin{equation}
{\cal L}_{WZ}^{(1)}\,=\,-\,{\frac{1}{2}}\,(\,\dot{\theta}^{2}\,+\,3\,{\theta
^{\prime }}^{2}\,)\,-\,\lambda \,(\,\dot{\theta}\,+\,\theta ^{\prime
}\,)\,-\,{\frac{1}{2}}\,\lambda ^{2}\;\;,
\end{equation}
where $\theta$ is the BF field, once integrated in the $\lambda $ field, a
chiral boson is recovered. Besides, if one takes the second WZ term
introduced in \cite{miao2}, 
\begin{equation}
{\cal L}_{WZ}^{(2)}\,=\,-\,\dot{\theta}\,\theta ^{\prime }\,-\,{\theta
^{\prime }}^{2}\,-\,\lambda \,(\,\dot{\theta}\,+\,\theta ^{\prime }\,)\,-\,{%
\frac{1}{2}}\,\lambda ^{2},
\end{equation}
and perform again the integration $\lambda $, one gets nothing but the
Lagrangian density of the free scalar boson. This result signalizes that the
WZ term obtained by Miao {\it et al} \cite{miao2} is already composed of two
opposite FJ's particles. Obviously it really introduces the degree of
freedom, because it is already there, in the WZ term, but it does not change
the physics of the SCB model, since, as we saw above, this last is composed
of vacuum.

Analyzing the interference aspects, we can apply the soldering formalism
again, but now, let us do it using two actions of the type of (\ref{final}),
i.e., 
\begin{eqnarray}
{\cal L}_1 &=& \dot{\eta}\,\eta^{\prime}\,-\,{\eta^{\prime}}^2\,-\,(\dot{%
\sigma}\,\sigma^{\prime}\,-\,{\sigma^{\prime}}^2) \\
{\cal L}_2 &=& -\,\dot{\xi}\,\xi^{\prime}\,-\,{\xi^{\prime}}^2\,-\,(-\,\dot{%
\omega}\,\omega^{\prime}\,-\,{\omega^{\prime}}^2)
\end{eqnarray}
where $\eta, \sigma, \xi$ and $\omega$ are all FJ's particles. We can see in
Eqs. (24) and (25) that the fields $(\eta,\xi)$ and $(\sigma,\omega)$ form
opposite chiralities particles pairs .

Performing the soldering procedure, one can easily see that the result is 
\begin{equation}
{\cal L}\,=\,{\frac{1}{2}}\,(\partial_{\mu}\,\Psi)\,-\,{\frac{1}{2}}%
\,(\partial_{\mu}\,\Lambda)
\end{equation}
where $\Psi=\eta-\xi$ and $\Lambda=\sigma-\omega$. This is the expected
result, and not (16), since we know that the SCB have a vacuum-like
spectrum. The soldering procedure in (26) discloses the same behavior as
shown in (\ref{final}).

The result (26) is quite different as the one shown in (16). Since it is
well known that the soldering of two opposite FJ chiral bosons is a massless
scalar field, we should expect that the fusion of two SCB would be two
opposite scalar fields with the final vacuum-like spectrum. This difference
can be explained \cite{aw} as we note that now, in each action of the Eqs.
(24) and (25), we have two fields, i.e., the action can be separated in two
different sectors, representing the FJ particles with the same chirality.
So, in the interference process (soldering) each sector of each action
interfere with its opposite partner. To obtain (16), note that we have only
one sector in each action. In the interference process we have lost the
information about the other sector, like a destructive interference. This
does not occur in (26).

\section{Conclusions}

It is well known that the SCB has consistency problems. In this work we have
used the soldering formalism to show that the interference on-shell of two
Srivastava's chiral bosons resulted in a scalar field. The other aspect of
this result is that the soldering method recover the consistency of the SCB
model, i.e., the fusion of opposites chiralities of the model results in a
consistent theory. This contradicts the conclusion published in the
literature, which asserts that it is impossible to apply the soldering
procedure to the SCB due to the inequivalence of this model with relation to
Siegel and Floreanini-Jackiw's models.

This has motivated us to explore the model promoting a canonical
transformation in the specific form of a dynamical decomposition, which
permitted us to decompose the action in its dynamical parts. This procedure
showed us that the SCB is in fact formed by two Floreanini-Jackiw's chiral
bosons of the same chiralities. Again, the contradiction with the current
literature is evident since one well known publication affirms that the WZ
term introduced a new degree of freedom to the theory resulting in two
Floreanini-Jackiw's chiral bosons of opposite chiralities, a chiral boson
and an antichiral boson. This is not really true, since we saw that in fact
the WZ term used consists of two degrees of freedom, i.e., the two FJ's
opposite chiral particles. So, one can say that it is obvious that this WZ
term should introduce new degrees of freedom because it is composed by the
fields that appeared. Now, with each SCB composed of two fields, after the
fusion through the soldering formalism, we have obtained two scalar fields
with a negative signal between them. This result shows that the spectrum of
the soldered action is vacuum-like.

\section{Acknowledgments}

The authors would like to thank C. Wotzasek and S. J. Gates Jr. for valuable
discussions. EMCA is financially supported by Funda\c{c}\~{a}o de Amparo
\`{a} Pesquisa do Estado de S\~{a}o Paulo (FAPESP). This work is partially
supported by Conselho Nacional de Pesquisa e Desenvolvimento (CNPq). FAPESP
and CNPq are brazilian research agencies.



\end{document}